# Tunable giant magnetoresistance in a single-molecule junction

Kai Yang[1†], Hui Chen[1†], Thomas Pope[2†], Yibin Hu[3†], Liwei Liu[1], Dongfei Wang[1], Lei Tao[1], Wende Xiao[1], Xiangmin Fei[1], Yu-Yang Zhang[1], Hong-Gang Luo[4], Shixuan Du[1], Tao Xiang[1], Werner A. Hofer[1,2]* and Hong-Jun Gao[1]*

[1]Institute of Physics & University of Chinese Academy of Sciences, Chinese Academy of Sciences, Beijing 100190, China
[2]School of Natural and Environmental Sciences, Newcastle University, Newcastle NE1 7RU, UK
[3]State Key Laboratory of Infrared Physics, Shanghai Institute of Technical Physics, Chinese Academy of Sciences, Shanghai 200083, China
[4]School of Physical Science and Technology, Lanzhou University, Lanzhou 730000, China

*Corresponding authors: H.-J. G. (hjgao@iphy.ac.cn) and W.A.H. (Werner.Hofer@newcastle.ac.uk)

†These authors contributed equally to this work.

**Controlling electronic transport through a single-molecule junction is crucial for molecular electronics or spintronics. In magnetic molecular devices, the spin degree-of-freedom can be used to this end since the magnetic properties of the magnetic ion centers fundamentally impact the transport through the molecules. Here we demonstrate that the electron pathway in a single-molecule device can be selected between two molecular orbitals by varying a magnetic field, giving rise to a tunable anisotropic magnetoresistance up to 93%. The unique tunability of the electron pathways is due to the magnetic reorientation of the transition metal center, resulting in a re-hybridization of molecular orbitals. We obtain the tunneling electron pathways by Kondo effect, which manifests either as a peak or a dip line shape. The energy changes of these spin-reorientations are remarkably low and less than one millielectronvolt. The large tunable anisotropic magnetoresistance could be used to control electronic transport in molecular spintronics.**

## Introduction

Single molecular magnets in contact with metal electrodes are exciting playgrounds to study spin-dependent transport through a single molecule[1-5], to examine fundamental magnetic interactions[6] and quantum many-body phenomena[3,4], and they are essential building blocks for spintronics devices[1,7,8]. The molecular spin states can be controlled mechanically[3], electrically[4], or by magnetic field[4,9,10], providing a useful handle to tune the flow of current. For example, using a magnetic field, spin-dependent transport phenomena can be tuned at the single-molecule level such as the negative differential resistance[9] or the highly-correlated Kondo effect[10].

Here, we measured the transport through a single iron phthalocyanine (FePc) molecule using a scanning tunneling microscope (STM), which provides an ideal experimental platform to study the single-molecule transport in an atomically well-defined environment[5,9,11-13]. Our single-molecule device consists of an FePc molecule attached to a Au(111) surface and an STM tip (Fig. 1). The two metal electrodes (Au and tip) serve as the source and drain. Normally, an electrostatic gate is used to shift the chemical potential of the molecular level[12,14]. In our experiment, we use an external magnetic field as a handle to control the spin states and the conductance of the single-molecule junction.

The FePc molecule belongs to a class of planar metal-phthalocyanine molecules, which have an indispensable role in spintronic applications due to the wide range of tunability of the spin-bearing centers[1,15-20]. The interaction of the magnetic molecules with a metal substrate can lead to a collective quantum behavior such as the Kondo effect in which the spin moment is screened by the coherent spin-flip process of the conduction electrons, giving rise to a Kondo resonance at the Fermi level ($E_F$)[10,21]. The line shape of Kondo resonance in the conductance spectra is sensitive to the local electronic and magnetic environment, such as conformational changes[22], inter-molecular interaction[11], charge donation[23], axial coordination[24,25], and filling of $d$-orbitals[16,23,26].

We use the line shape and spatial distribution of the Kondo resonance of FePc as indicators to monitor the electron pathway in the tip-FePc-Au junction and demonstrate that the electrons travel through the FePc by two possible $d$ orbitals: $d_{z^2}$ and $d_\pi$ ($d_\pi$ represents $d_{xz}$ or $d_{yz}$), and that the single-electron passage through the FePc can be controlled by the external magnetic field (Fig. 1). Density functional theory (DFT) calculations reveal that this unique tunability originates from the reorientation of the magnetic moment on the Fe atom.

## Results

**Topographic image of FePc.** After adsorption on Au(111), the FePc molecule appears in STM images as a cross with a central protrusion (Fig. 2a). The FePc molecule under study adsorbs at the bridge site of Au(111) (Supplementary Fig. 1). Our DFT calculation shows that the total spin magnetic moment of Fe is 2.05 $\mu_B$ and, in the ground state, it orients close to the plane of the molecule (easy-plane magnetic anisotropy).

**Temperature and magnetic-field dependence of dI/dV spectra.** To resolve the molecular spin states, we measured differential conductance (d$I$/d$V$) spectra at the Fe center of the molecule. At zero magnetic field, the d$I$/d$V$ spectrum exhibits a sharp dip around $E_F$ (Fig. 2a). The dip intensity decreases rapidly with increasing temperature and disappears completely above 8 K (Fig. 2b). The sharp dip at $E_F$ is attributed to the Kondo effect since the dip conductance decreases logarithmically with decreasing temperature (Supplementary Fig. 2a), which is a characteristic feature of the Kondo effect[27].

Figure 2c shows the evolution of the d$I$/d$V$ spectra at different magnetic field. The sharp Kondo dip at $E_F$ becomes shallower but its width remains almost unchanged with increasing magnetic field for $B < 4$ T. At $B = 4$ T the Kondo dip is completely suppressed. When the magnetic field is above 4 T, a peak emerges at $E_F$ and becomes more pronounced and broader with increasing magnetic field.

**Spatial distribution of the Kondo resonance.** The dip-to-peak transition of the Kondo resonance indicates the change of the molecular spin states with magnetic field. To gain a better understanding of the spin states, we studied the spatial distribution of the Kondo resonance over the FePc molecule by performing spectroscopic mapping near $E_F$ under varying magnetic fields (Fig. 3 and Supplementary Fig. 3). At zero or low fields, the Kondo resonance shows a non-circular spatial distribution (Fig. 3a). This distribution is not due to drifting during measurement, as confirmed by d$I$/d$V$ spectra taken along different axes over the FePc molecule (Figs. 3b and 3c). It shows that the Kondo resonance (dip) decays slower along one axis of the feature (Fig. 3c) than the other (Fig. 3b). In comparison, at high magnetic fields (above 4 T), the Kondo resonance (peak) is more localized on the molecular center, with a radial symmetry (Figs. 3d-f). At intermediate field strengths of 3 T to 6 T, the Kondo resonance is largely invisible (Supplementary Figs. 3c and d).

The spatial distribution of the Kondo resonance is determined by the electron orbital responsible for the resonance, that is, the local distribution of electron charge of the responsible orbital[11,28]. The two different spatial distributions (non-circular or circular) suggest two distinct Kondo screening channels originating from two $d$ orbitals at different magnetic fields. The change of spatial distributions of Kondo resonance can be understood by considering the different spatial symmetries of the $d_\pi$ and $d_{z^2}$ orbitals (Fig. 1). The hybridization of $d_\pi$ orbital with the neighbouring atoms inside the FePc molecule results in a more extended Kondo resonance distribution at low field; while the $d_z^2$ orbital, due to its different symmetry, hybridizes less with the molecular orbitals, giving rise to the more localized spatial distribution of Kondo resonance at high magnetic field.

The change of orbital characteristics results from the reorientation of the Fe magnetic moment induced by the applied magnetic field. This reorientation changes the orbital character through spin-orbit interaction[29]. The applied magnetic field flips the magnetic moment of Fe from the in-plane to the out-of-plane direction when the Zeeman energy is larger than the magnetic anisotropy energy. Thus, due to the spin–orbit interaction, the reorientation of the magnetic moment changes the mixing of different $d$ orbital character in the ground states. A level crossing within the low-energy spin multiplet with quenched orbital angular momentum cannot explain the change of the spatial distribution of the Kondo resonance with increasing magnetic field, since this scenario would only result in a tiny change of the orbital composition in the wave function.

**Reorientation of the Fe magnetic moment.** To confirm this, we performed noncollinear DFT calculations for both the in-plane configuration (Fe magnetic moment in the plane of the molecule), and the out-of-plane configuration (Fe magnetic moment aligned to the magnetic field, perpendicular to the molecular plane). Our DFT calculations show that at zero magnetic field the Fe magnetic moment is in the plane of the molecule. The reorientation of the magnetic moment at strong field is possible because the Zeeman energy due to the external magnetic field (~ 4 T) is comparable with the in-plane magnetic anisotropy of the Fe spin (~ 1 meV from DFT calculations). Similar energy scale of this magnetic anisotropy energy was obtained by other simulations[30]. Figures 4a and 4b show the calculated band-decomposed partial density of states (PDOS) of the $d$ orbitals for both magnetic configurations. By comparing spatial distribution of the molecular orbitals closest to $E_F$ (Figs. 4c, and 4d) with the experimental spatial distribution of the Kondo resonance, we find that the non-circular depression at low field is due to the $d_\pi$ orbital (Fig. 4c), while the circular protrusion at high field is due to the $d_z^2$ orbital (Fig. 4d). We also find that the relative population of the two orbitals changes as the magnetic moment orientates onto the direction of the magnetic field. The re-hybridization of the molecular orbitals of the magnetic ground state thus gives rise to the change of spatial distribution of the zero-bias conductance (Kondo resonance) over the FePc molecule as shown in Fig. 3. Note that the reorientation of the magnetic moment doesn't mean that the Zeeman energy induces direct orbital transitions, since only the low-energy density distribution of the $d$ orbitals near $E_F$ changes.

The re-hybridization in the ground state of FePc can also be directly visualized in the change of the line shapes of the Kondo resonance, in addition to its spatial distribution. The dip-to-peak transition of the Kondo resonance with external magnetic field (Fig. 2c) can be understood by considering the different spatial symmetries of the $d$ orbitals. The line shapes of the Kondo resonance are determined by the Fano interference during the electron tunneling process[31]. Electrons originating from the tip either tunnel through the discrete Kondo resonance or the continuum states of the conduction electrons. The Fano asymmetry factor $q$ is proportional to the ratio of the probabilities of the two paths. Since different $d$ orbitals couple differently to the tip due to their alignments and shapes, this will result in different Fano interference with the tunneling into the Au state as the magnetic field increases. At high magnetic field, the magnetic ground state is mainly contributed by the $d_{z^2}$ orbital (Fig. 4d), which couples more to the tip states due to its favorable spatial distribution (Fig. 1b). Hence at high field the Kondo resonance appears as a peak (large $q$-factor). Similarly, the weaker coupling between the $d_\pi$ and the tip states results in the Kondo dip at low magnetic fields (Fig. 1a).

We fit the Kondo resonances at different magnetic field with a Fano function (Supplementary Fig. 4)[21]. The fitting shows that the Fano asymmetry factor $q$ drops suddenly from positive to negative values across the transition magnetic field (4 T), corresponding to the dip-to-peak spectral transition. The line width of the Kondo resonance increases with magnetic field above 6 T due to Zeeman splitting[27,32]. This sudden change in the spectral feature – going from a configuration in which it is unaffected by Zeeman splitting into a configuration in which Zeeman splitting is measured – supports the argument that the magnetic moment is reoriented. In the in-plane configuration, the magnetic moment is nearly perpendicular to the magnetic field and thus the product is small. At high field, the reorientation allows for a much larger product and, so, noticeable Zeeman splitting (Supplementary Fig. 5).

**Tunneling anisotropic magnetoresistance of FePc.** The magnetic reorientation of the Fe center by varying the magnetic field also leads to the tunneling anisotropic magnetoresistance (TAMR) effect. The TAMR effect describes the dependence of the magnetoresistance on the magnetization orientation, and has been found in metal film[33,34], single adatoms[29,35], and molecular tunneling junctions[36-39]. In addition, the TAMR effect with non-magnetic electrodes has been shown in mechanical break junctions[40,41]. Here, the magnetic FePc molecule is attached directly to a gold metal electrode in our tunnel junction and there is no additional magnetic layer.

The TAMR can be defined as TAMR($M$) = (d$I$(0)/d$V$ – d$I$($M$)/d$V$)/(d$I$(0)/d$V$), where d$I$($M$)/d$V$ denotes the differential conductance of FePc at the magnetic reorientation with an out-of-plane component $M$[39]. Increasing the magnetic field corresponds to a change of the magnetization angle from zero to 90 degree. We show the TAMR of FePc as a function of bias voltage in Fig. 4e. In the Kondo resonance region (between ±5 mV), the TAMR can be enhanced up to 93%, much larger than the TAMR values reported in other molecular-based TAMR devices[36-39]. At intermediate magnetic field (Fig. 4f), the TAMR effect (with respect to the in-plane magnetization at zero field) increases monotonically with the magnetic field near zero bias.

## Discussion

To demonstrate the generality of tuning the tunnelling pathway with magnetic fields, we also studied the Kondo effect of an Fe-porphyrin derivative on Au(111). The Kondo resonance there shows similar dip-to-peak transitions with magnetic fields (Supplementary Fig. 6d). We expect the behavior of the magnetic moment to be replicated in the Fe-porphyrin derivative. The d$I$/d$V$ mapping of the Kondo resonance exhibits a similar transition from an extended distribution to a more concentrated distribution (Supplementary Figs. 6e-h).

We can treat the electron tunneling through the FePc molecule as single-electron transport. The time lag from one electron to the next through the molecule is in the range of nanoseconds (corresponding to a tunneling current of nanoamperes), but other processes at the atomic scale occur at a much faster time scale. The fastest processes, electronic relaxations, typically occur within femtoseconds. The dynamics of the atomic cores, manifest by phonons and vibrations, are within the picosecond range, with relaxation times such that they typically do not exceed a few hundred picoseconds. The transport phenomena in the STM junction are thus too slow for vibrations or electronic relaxations to influence electron transport.

In conclusion, we have demonstrated that the pathway of single electrons through the orbitals of a magnetic molecule can be tuned at a very low energy by varying the external magnetic field. This unique tunability originates from the reorientation of the magnetic moment on the metal center, which alters the electron distribution in the $d$ orbitals. While at zero field the electron density scattered through the Kondo resonance will be exclusively at the $d_\pi$ orbital, it will pass through the $d_{z^2}$ orbital at high field, and a varying part of the density will pass through both orbitals in the intermediate regime. Note that a local control of the spin reorientation could in principle be achieved by using the magnetic field from a spin-polarized STM tip[42]. Our work shows that the multi-orbital nature and the spin-orbit coupling can be employed to control the single-electron process in a single-molecule device at sub-meV energies.

## Methods

**Experiment.** The atomically flat Au(111) surface was prepared by repeated cycles of sputtering with argon ions and annealing at 800 K. Commercial FePc molecules (Sigma-Aldrich, 97% purity) were sublimated from a Knudsen-type evaporator after thermal purification, while the Au(111) substrate was held at room temperature. STM images were acquired in the constant-current mode and all given voltages are referred to the sample. d$I$/d$V$ spectra were measured using a lock-in technique with a sinusoidal modulation (0.09 mV rms ~ 0.3 mV rms) at a frequency of 973 Hz. The electrochemically etched tungsten tips were calibrated against the Au(111) surface state before spectroscopic measurements to ensure no tip-related features on the recorded d$I$/d$V$ spectra.

**Ab-initio Calculations.** All non-collinear spin-polarized density functional theory (DFT) calculations were performed with the Vienna ab-initio Simulation Package (VASP)[43-46] and the projector augmented wave (PAW) method[47,48]. The exchange-correlation potential was described by the Perdew-Burke-Ernzerhof functional[49] extended to incorporate a Van der Waals correction[50]. We perform a geometry minimization on the system, until the residual forces were smaller than 0.02 eV Ang$^{-1}$. To improve the description of band structure and magnetic properties of the adsorbed FePc molecule, a Hubbard-like +$U$ correction is adopted[51,52]. The Hubbard-like on-site Coulomb and exchange parameters for the Fe 3$d$ orbitals are respectively chosen to be $U = 2$ eV and $J = 1$ eV. These values have been used in previous studies on the same system[53-55]. In calculating the density of states, we employed Gaussian smearing with a width of 5 meV.

## Data availability

The data that support the findings of this study are available from the corresponding authors upon reasonable request.

# References

1. L. Bogani & W. Wernsdorfer. Molecular spintronics using single-molecule magnets. *Nat. Mater.* **7**, 179-186 (2008).

2. W. Liang *et al.* Kondo resonance in a single-molecule transistor. *Nature* **417**, 725-729 (2002).

3. J. J. Parks *et al.* Mechanical control of spin states in spin-1 molecules and the underscreened Kondo effect. *Science* **328**, 1370-1373 (2010).

4. N. Roch *et al.* Quantum phase transition in a single-molecule quantum dot. *Nature* **453**, 633-637 (2008).

5. S. Schmaus *et al.* Giant magnetoresistance through a single molecule. *Nat. Nanotechnol.* **6**, 185-189 (2011).

6. C. F. Hirjibehedin, C. P. Lutz & A. J. Heinrich. Spin coupling in engineered atomic structures. *Science* **312**, 1021-1024 (2006).

7. S. Sanvito. Molecular spintronics. *Chem. Soc. Rev.* **40**, 3336-3355 (2011).

8. S. V. Aradhya & L. Venkataraman. Single-molecule junctions beyond electronic transport. *Nat. Nanotechnol.* **8**, 399-410 (2013).

9. B. Warner *et al.* Tunable magnetoresistance in an asymmetrically coupled single-molecule junction. *Nat. Nanotechnol.* **10**, 259-263 (2015).

10. G. D. Scott & D. Natelson. Kondo resonances in molecular devices. *ACS Nano* **4**, 3560-3579 (2010).

11. A. Mugarza *et al.* Spin coupling and relaxation inside molecule–metal contacts. *Nat. Commun.* **2**, 490 (2011).

12. J. Martínez-Blanco *et al.* Gating a single-molecule transistor with individual atoms. *Nat. Phys.* **11**, 640-644 (2015).

13. T. Esat, N. Friedrich, F. S. Tautz & R. Temirov. A standing molecule as a single-electron field emitter. *Nature* **558**, 573-576 (2018).

14. I. Fernández-Torrente *et al.* Gating the charge state of single molecules by local electric fields. *Phys. Rev. Lett.* **108**, 036801 (2012).

15. L. Gao *et al.* Site-specific Kondo effect at ambient temperatures in iron-based molecules. *Phys. Rev. Lett.* **99**, 106402 (2007).

16. S. Stepanow *et al.* Mixed-valence behavior and strong correlation effects of metal phthalocyanines adsorbed on metals. *Phys. Rev. B* **83**, 220401 (2011).

17. N. Tsukahara *et al.* Evolution of Kondo resonance from a single impurity molecule to the two-dimensional lattice. *Phys. Rev. Lett.* **106**, 187201 (2011).

18	N. Ballav *et al.* Emergence of on-surface magnetochemistry. *J. Phys. Chem. Lett.* **4**, 2303-2311 (2013).

19	W. Auwarter, D. Ecija, F. Klappenberger & J. V. Barth. Porphyrins at interfaces. *Nat. Chem.* **7**, 105-120 (2015).

20	T. Pope, S. Du, H.-J. Gao & W. A. Hofer. Electronic effects and fundamental physics studied in molecular interfaces. *Chem. Commun.* **54**, 5508-5517 (2018).

21	V. Madhavan *et al.* Tunneling into a single magnetic atom: Spectroscopic evidence of the Kondo resonance. *Science* **280**, 567-569 (1998).

22	T. Komeda *et al.* Observation and electric current control of a local spin in a single-molecule magnet. *Nat. Commun.* **2**, 217 (2011).

23	C. Krull, R. Robles, A. Mugarza & P. Gambardella. Site- and orbital-dependent charge donation and spin manipulation in electron-doped metal phthalocyanines. *Nat. Mater.* **12**, 337-343 (2013).

24	A. Stróżecka, M. Soriano, J. I. Pascual & J. J. Palacios. Reversible change of the spin state in a manganese phthalocyanine by coordination of CO molecule. *Phys. Rev. Lett.* **109**, 147202 (2012).

25	L. W. Liu *et al.* Reversible single spin control of individual magnetic molecule by hydrogen atom adsorption. *Sci. Rep.* **3**, 1210 (2013).

26	J. Kügel *et al.* Relevance of hybridization and filling of 3d orbitals for the Kondo effect in transition metal phthalocyanines. *Nano Lett.* **14**, 3895-3902 (2014).

27	S. M. Cronenwett, T. H. Oosterkamp & L. P. Kouwenhoven. A tunable Kondo effect in quantum dots. *Science* **281**, 540-544 (1998).

28	G. E. Pacchioni *et al.* Two-orbital Kondo screening in a self-assembled metal–organic complex. *ACS Nano* **11**, 2675-2681 (2017).

29	N. Néel *et al.* Tunneling anisotropic magnetoresistance at the single-atom limit. *Phys. Rev. Lett.* **110**, 037202 (2013).

30	J. Hu & R. Wu. Control of the magnetism and magnetic anisotropy of a single-molecule magnet with an electric field. *Phys. Rev. Lett.* **110**, 097202 (2013).

31	T. Markus, J. H. Andreas & S. Wolf-Dieter. Spectroscopic manifestations of the Kondo effect on single adatoms. *J. Phys.: Condens. Matter* **21**, 053001 (2009).

32	D. Goldhaber-Gordon *et al.* Kondo effect in a single-electron transistor. *Nature* **391**, 156-159 (1998).

33	M. Bode *et al.* Magnetization-direction-dependent local electronic structure probed by scanning tunneling spectroscopy. *Phys. Rev. Lett.* **89**, 237205 (2002).

34	M. Hervé *et al.* Stabilizing spin spirals and isolated skyrmions at low magnetic field exploiting vanishing magnetic anisotropy. *Nat. Commun.* **9**, 1015 (2018).

35   D. Serrate *et al.* Imaging and manipulating the spin direction of individual atoms. *Nat. Nanotechnol.* **5**, 350-353 (2010).

36   M. Grünewald, N. Homonnay, J. Kleinlein & G. Schmidt. Voltage-controlled oxide barriers in organic/hybrid spin valves based on tunneling anisotropic magnetoresistance. *Phys. Rev. B* **90**, 205208 (2014).

37   C. Barraud *et al.* Unidirectional spin-dependent molecule-ferromagnet hybridized states anisotropy in cobalt phthalocyanine based magnetic tunnel junctions. *Phys. Rev. Lett.* **114**, 206603 (2015).

38   T. Kamiya, C. Miyahara & H. Tada. Large tunneling anisotropic magnetoresistance in $La_{0.7}Sr_{0.3}MnO_3$/pentacene/Cu structures prepared on $SrTiO_3$ (110) substrates. *Appl. Phys. Lett.* **110**, 032401 (2017).

39   K. Wang *et al.* Effect of orbital hybridization on spin-polarized tunneling across $Co/C_{60}$ interfaces. *ACS Appl. Mater. Interfaces* **8**, 28349-28356 (2016).

40   R. Hayakawa *et al.* Large magnetoresistance in single-radical molecular junctions. *Nano Lett.* **16**, 4960-4967 (2016).

41   Z. Xie *et al.* Large magnetoresistance at room temperature in organic molecular tunnel junctions with nonmagnetic electrodes. *ACS Nano* **10**, 8571-8577 (2016).

42   S. Yan *et al.* Control of quantum magnets by atomic exchange bias. *Nat. Nanotechnol.* **10**, 40-45 (2015).

43   G. Kresse & J. Hafner. Ab initio molecular dynamics for liquid metals. *Phys. Rev. B* **47**, 558-561 (1993).

44   G. Kresse & J. Furthmüller. Efficiency of ab-initio total energy calculations for metals and semiconductors using a plane-wave basis set. *Comp. Mater. Sci.* **6**, 15-50 (1996).

45   G. Kresse & J. Furthmüller. Efficient iterative schemes for ab initio total-energy calculations using a plane-wave basis set. *Phys. Rev. B* **54**, 11169-11186 (1996).

46   D. Hobbs, G. Kresse & J. Hafner. Fully unconstrained noncollinear magnetism within the projector augmented-wave method. *Phys. Rev. B* **62**, 11556-11570 (2000).

47   P. E. Blöchl. Projector augmented-wave method. *Phys. Rev. B* **50**, 17953-17979 (1994).

48   G. Kresse & D. Joubert. From ultrasoft pseudopotentials to the projector augmented-wave method. *Phys. Rev. B* **59**, 1758-1775 (1999).

49   J. P. Perdew, K. Burke & M. Ernzerhof. Generalized gradient approximation made simple. *Phys. Rev. Lett.* **77**, 3865-3868 (1996).

50   S. Grimme. Semiempirical GGA-type density functional constructed with a long-range dispersion correction. *J. Comput. Chem.* **27**, 1787-1799 (2006).

51  A. I. Liechtenstein, V. I. Anisimov & J. Zaanen. Density-functional theory and strong interactions: Orbital ordering in Mott-Hubbard insulators. *Phys. Rev. B* **52**, R5467-R5470 (1995).

52  S. L. Dudarev *et al*. Electron-energy-loss spectra and the structural stability of nickel oxide: An LSDA+U study. *Phys. Rev. B* **57**, 1505-1509 (1998).

53  J. M. Gottfried. Surface chemistry of porphyrins and phthalocyanines. *Surf. Sci. Rep.* **70**, 259-379 (2015).

54  Y. Wang, X. Zheng & J. Yang. Environment-modulated Kondo phenomena in FePc/Au(111) adsorption systems. *Phys. Rev. B* **93**, 125114 (2016).

55  Y. Wang, X. Li, X. Zheng & J. Yang. Spin switch in iron phthalocyanine on Au(111) surface by hydrogen adsorption. *J. Phys. Chem.* **147**, 134701 (2017).

## Acknowledgements

Work at IOP was supported by the Strategic Priority Research Program of Chinese Academy of Sciences (XDB30000000), the International Partnership Program of Chinese Academy of Sciences (112111KYSB20160061), National Key Research and Development Projects of China (2016YFA0202300, 2017YFA0302900), and the National Natural Science Foundation of China (61888102). W.A.H. acknowledges support for the North East Centre for Energy Materials, EPSRC Grant EP/R021503/1. This research made use of the Rocket High Performance Computing service at Newcastle University.

## Author contributions

K.Y., H.C., T.P. and Y.H. contributed equally to this work. K.Y., H.C., L.W.L., D.F.W., W.D.X. and X.M.F. performed STM measurements. T.P., Y.H., L.T., Y.Y.Z., H.G.L., S.X.D., T.X. and W.A.H. carried out theoretical calculations. K.Y., H.C., T.P., Y.H., W.D.X., S.X.D., W.A.H. and H.-J.G. analyzed the data and wrote the manuscript. H.-J.G. designed and coordinated the project.

## Competing interests

The authors declare no competing interests.

# Figures

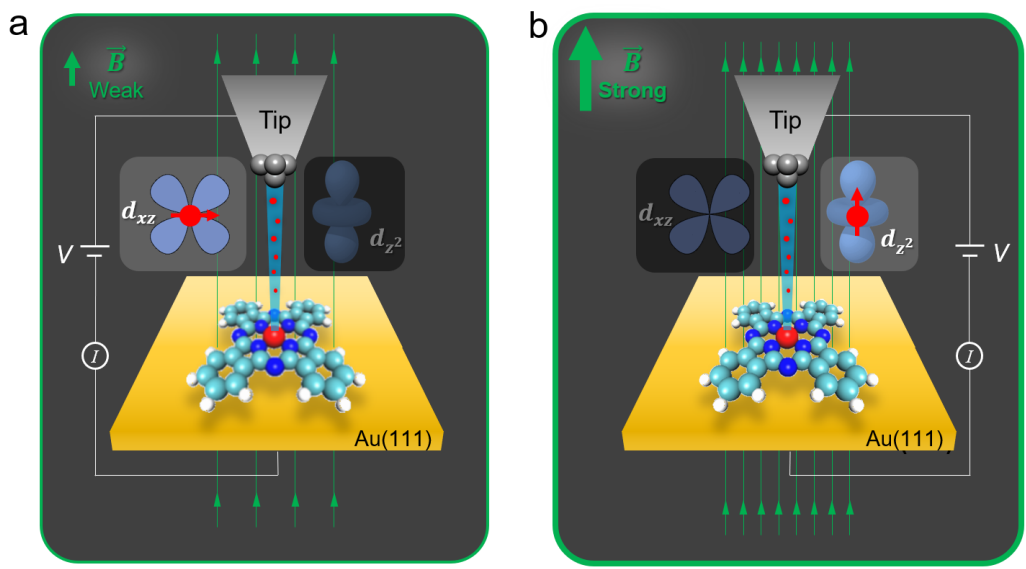

**Figure 1 | A tunable single-molecule device in the STM junction. a** and **b**, Schematics of the electron transport process through an FePc molecule adsorbed on a Au(111) surface at different magnetic fields. The Au substrate and STM tip are the two terminals of the single-molecule device. The magnetic field can be viewed as a gate to control the molecular spin states. The arrays of green lines indicate the magnetic field lines. During the electron transport through the FePc, the tunneling electron has two possible passages corresponding to two molecular orbitals ($d_z^2$ or $d_{xz}/d_{yz}$). At weak magnetic field as in **a**, the spin direction of Fe is in plane and currents flow by the $d_{xz}/d_{yz}$ orbital. At strong magnetic field as in **b**, Fe spin is aligned to the magnetic field and electrons tunnel preferentially through the $d_z^2$ orbital.

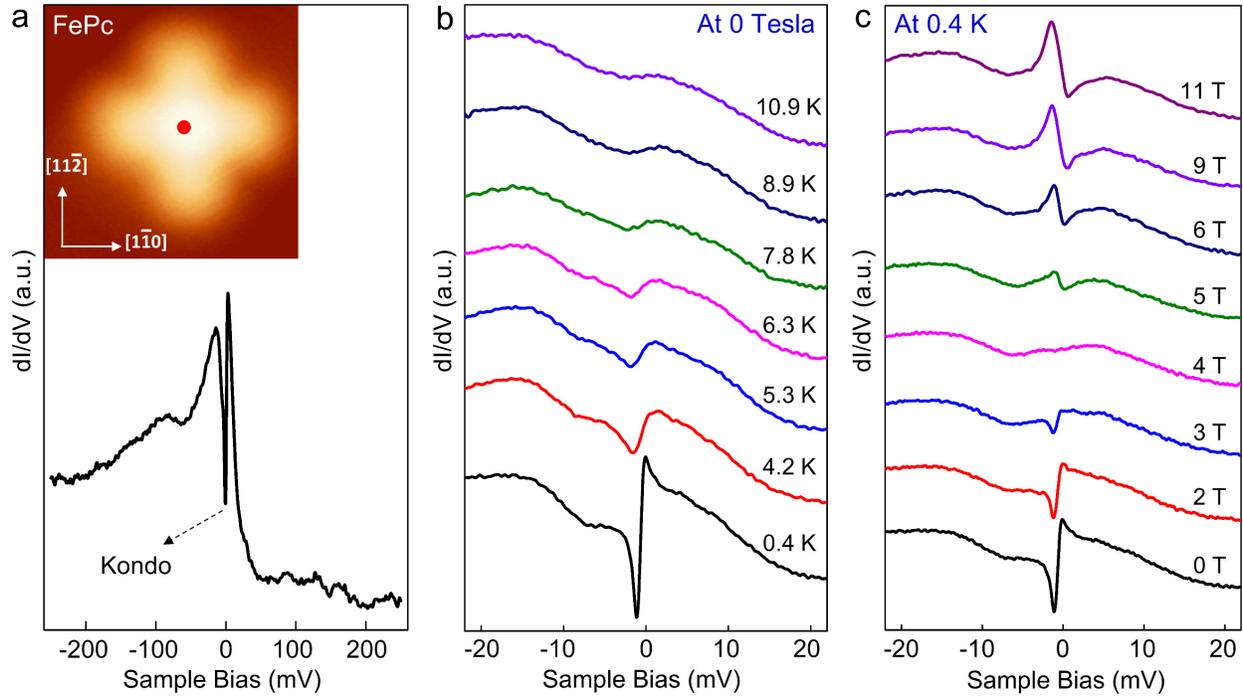

**Figure 2 | d$I$/d$V$ spectra under variation of temperature and magnetic field. a,** d$I$/d$V$ spectra taken on the Fe center at 0.4 K and zero magnetic field, showing a Kondo dip superimposed on a broad feature near $E_F$ (setpoint: $I$ = 0.2 nA, $V_b$ = −0.1 V). Inset: STM image of an FePc on Au(111) (3.2 nm × 3.2 nm, $I$ = 10 pA, $V_b$ = −0.2 V). **b,** Evolution of the d$I$/d$V$ spectra of FePc at increasing temperatures in the absence of magnetic field (setpoint: $I$ = 0.3 nA, $V_b$ = −60 mV). Successive spectra are offset for clarity. The dip at $E_F$ vanishes above 8 K, making two broad resonances visible. The broad two-peak feature was originally attributed to a Kondo feature in Ref. [15]. However, in this work we note the similarity between the spectra and the density of states for the underlying gold bulk (Supplementary Fig. 7), suggesting that this is in fact the source of the broader feature. **c,** Evolution of the d$I$/d$V$ spectra of FePc with increasing magnetic field at 0.4 K (setpoint: $I$ = 0.3 nA, $V_b$ = −60 mV), showing a dip-to-peak transition. a.u., arbitrary units.

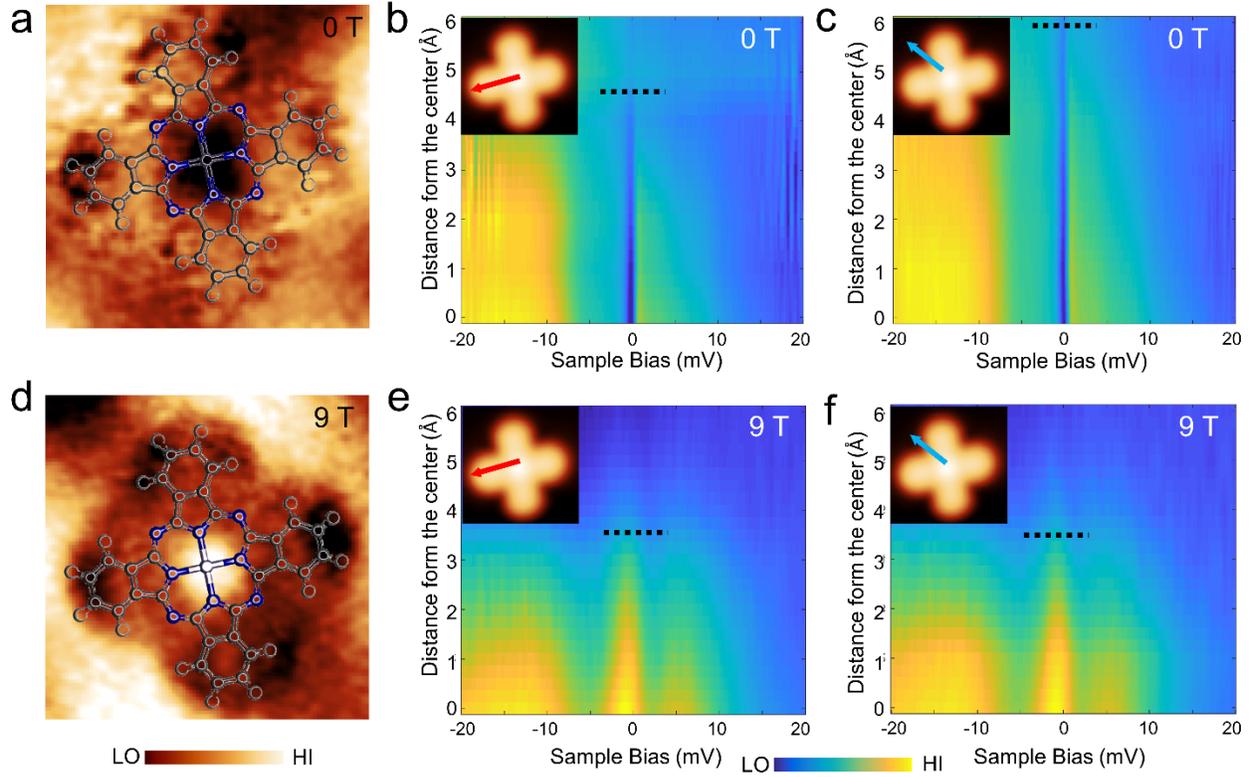

**Figure 3 | Spatial distribution of the Kondo resonance at different magnetic fields. a,** d$I$/d$V$ mapping taken around the Fermi energy at 0 T (setpoint: $I$ = 0.3 nA, $V_b$ = −40 mV. Image size: 2.2 nm × 2.2 nm). **b** and **c,** d$I$/d$V$ spectra taken along different directions on the FePc molecule (as indicated by the arrows). The spectra start from the Fe center and end at about 6 Å away. The black dashed line roughly indicates the position where the Kondo resonance is too weak to see in the d$I$/d$V$ spectra. Inset: STM image (2.2 nm × 2.2 nm) of the FePc. **d-f,** same as **a-c** except taken at $B$ = 9 T. The Kondo resonance (peak) displays a nearly circular spatial distribution in **d**.

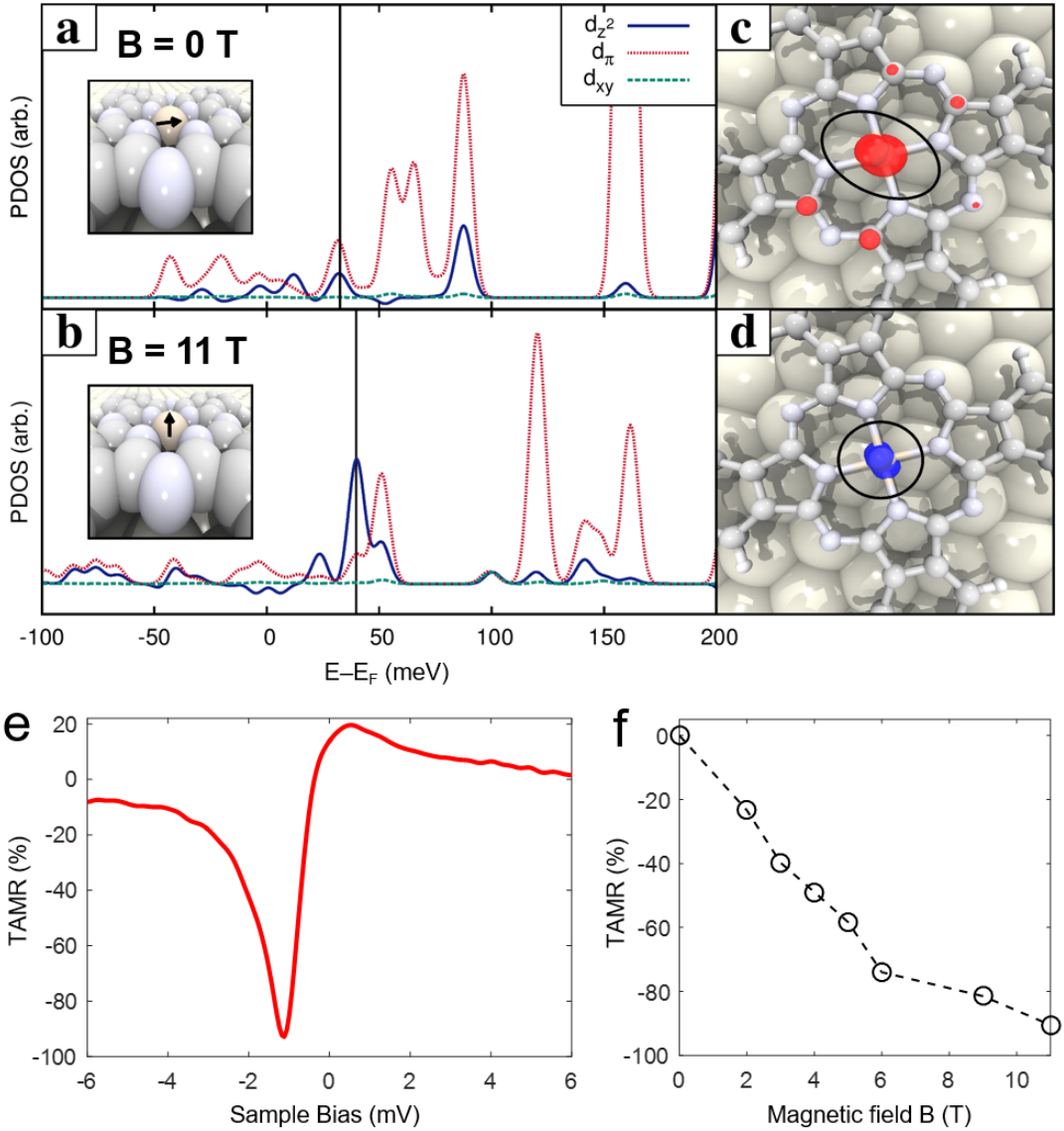

**Figure 4 | Controlling the tunneling anisotropic magnetoresistance (TAMR). a** and **b**, Calculated *lm*-decomposed partial density of states (PDOS) of the $d_{xy}$, $d_\pi$ and $d_{z^2}$ bands of the Fe atom with a magnetization vector in the plane of the molecule **a** and in the direction of the applied magnetic field, perpendicular to the plane of the molecule **b**. The insets show the direction of the magnetic moment schematically. The energy is given with respect to $E_F$. Black lines represent the energy of the molecular orbital closest to $E_F$. **c** and **d**, Band-decomposed charge density plots for the molecular orbitals close to $E_F$, showing the similar spatial distribution of the Kondo resonance as in Figs. 3a and 3d. We note that the orbital density on the C atoms in **c** is unlikely to have a significant overlap with the substrate and thus will not contribute significantly to the signals in the d$I$/d$V$ maps. **e,** TAMR values at different bias voltages calculated from the d$I$/d$V$ spectra taken on the Fe center at $B = 0$ and 11 T. **f,** TAMR values at different magnetic field calculated from the d$I$/d$V$ spectra taken on the Fe center at the bias voltage of −1.06 mV. a.u., arbitrary units.